\documentclass[9pt,twoside]{pnas-new}
\usepackage{bm}
\templatetype{pnasmathematics} 

\title{Hopfield-like open channel retrieval for disordered optical media}

\author[a,b,1]{Marco Leonetti} 
\author[a,c]{ Luca Leuzzi} 
\author[b,c]{Giancarlo Ruocco}

\affil[a]{Soft and Living Matter Laboratory, Institute of Nanotechnology, Consiglio Nazionale delle Ricerche, I-00185 Rome, Italy;}
\affil[b]{Center for Life Nano Science@Sapienza, Istituto Italiano di Tecnologia, I-00161 Rome, Italy}
\affil[c]{Department of Physics, University Sapienza, I-00185 Roma, Italy;}
\affil[d]{Sezione di Roma I, Istituto Nazionale di Fisica Nucleare, I-00185 Roma, Italy}

\leadauthor{Leonetti} 

\significancestatement{Wavefront shaping is a sophisticated optical technique based on multi-element adaptive optics devices, enabling to focus and shape light at the back of strongly scattering medium. The two most popular approaches (sequential optimization and transmission matrix measurement) are affected by limitations. Here we demonstrated that these limitations are due to the roughness of the intensity landscape and proposed new approach which is based on the same mathematics supporting efficient memory retrieval in Hopfield networks. Our Complete Couplings Mapping Method surpasses the problems connected to the Intensity landscape shape and can be implemented using a digital micro-mirror device, thus enabling fast laser scanning at the back of a disordered medium.}

\authorcontributions{Please provide details of author contributions here.}
\authordeclaration{Authors declare no competing interests}
\equalauthors{\textsuperscript{1}M.L.designed and performed the experiments G.R.and  L.L. developed the theory and analytical predictions. All the authors contributed to the realization of the manuscript }
\correspondingauthor{\textsuperscript{2}To whom correspondence should be addressed. E-mail: marco.leonetti\@gmail.com}


\begin{abstract} 
The measurement of the optical Transmission Matrix (TM) enables to access "open channels": input patterns, specific to each scattering structure, capable to deliver very high transmission. Various approaches, based either on multiple interferometric measurements or on systematic random testing of incident wavefronts, enable to estimate the inputs required to excite these open channels.

Here, we provide for the first time an approach enabling the complete and reference-less retrieval of the open channels. It is based on the full mapping all the pairwise interference terms resulting from all the input modes couples. We show that these interference terms are organized into a bi-dyadic coupling matrix whose eigenvalues enables to access the open channel. A disordered optical system, is thus behaving exactly like  an Hopfield neural network, where a specific input vector (an eigenvalue of the neurons' coupling matrix) enables to retrieve a specific memory pattern \cite{ramsauer2020hopfield,sabahi2016unsupervised}.
The proposed Hopfield like open-channel-retrieval  approach, enables to reach almost 100$\%$ of the theoretically expected value of the Intensity. Moreover employing a digital micromirror device to modulate light, we demonstrate high speed laser scanning at the back of a disordered medium.
\end{abstract}

\dates{This manuscript was compiled on \today}
\doi{\url{www.pnas.org/cgi/doi/10.1073/pnas.XXXXXXXXXX}}

\begin{document}

\maketitle
\thispagestyle{firststyle}
\ifthenelse{\boolean{shortarticle}}{\ifthenelse{\boolean{singlecolumn}}{\abscontentformatted}{\abscontent}}{}

\dropcap{T}he interaction between electromagnetic waves and opaque non-absorbing systems, may be modeled with the transmission matrix (TM) approach \cite{beenakker1997random, vanrossum1999multiple,vellekoop2007focusing,van2010information,PhysRevLett.104.100601,popoff2010image, popoff2014coherent,goorden2014superpixel,ploschner2015seeing,dremeau2015reference, del2016intensity,RevModPhys.89.015005,  zhao2020seeing,devaud2020speckle, li2021compressively,li2021memory,huang2021generalizing}: a  set of complex parameters describing attenuation and dephasing  with respect to the input. TM counter-intuitively predicts that an optimally prepared coherent input beam could be transmitted in extremely efficient fashion even trough extremely thick opaque optical media. This highly transmitting states are known as  \textbf{``open channels''} or  \textbf{``transmission eigenchannels''}  \cite{popoff2014coherent} and are linked to the eigenvectors of the disordered, complex transmission matrix.
In  general, open channels are found in two ways: either by a sequential intensity optimization algorithm (see the seminal paper of  Vellekoop and Mosk \cite{vellekoop2007focusing}) or by  transmission matrix phase measurement and  conjugation (as in an impressive series of works of Gigan and collaborators, see e. g., \cite{PhysRevLett.104.100601}). If we term $ I_{\rm focus}$ the maximum realizable intensity at a given focus  at the output  and $I_{\rm speckle}$ the intensity distributed in a generic random way in speckles at the output of the random medium, 
the  parameter defining the efficiency of the wavefront shaping technique is  $$\eta=\frac{I_{\rm focus}}{\langle I_{\rm speckle}\rangle}$$ 
where $\langle \rangle$ represent averaging over input configurations.
This, measures the effectiveness of a modulation technique to deliver light at a desired location with respect to the average speckle intensity obtained for non-optimal transmission. 
Regardless of the method employed to reconstruct the Transmission Matrix (TM), the maximum  achievable efficiency $\eta$ is connected to the number of total controlled segments $N$ in the input as \cite{nam2020increasing} 
\begin{equation}
    \eta=1 + \alpha(N-1)
    \label{eq:efficiency}
\end{equation}
where $\alpha$ is a factor - termed efficiency parameter - depending on the nature of the modulation. For instance, for full  phase modulation \cite{PhysRevLett.104.100601,yoon2015measuring} it is  $\alpha=\pi/4$, whereas $\alpha=1/(2 \pi)$ for binary amplitude modulation \cite{nam2020increasing, vellekoop2008universal,conkey2012high} and, as we will consider in this work,  $\alpha=1/\pi$ for binary phase modulation (see supplementary information file, S.I.). 

Here, first we will demonstrate that SO is not always capable to reach the maximal efficiency for light focusing and why this comes about, then we will provide a new method to focus light behind a disordered medium which is reference less. We will demonstrate for the first time that this method  is capable to return an intensity enhancement with is extremely close to the theoretical predictions.  We will name the approach the  Complete Couplings Mapping Method.

We start defining the problem. The  field $E^{(\nu)}$ at a given target location $\nu$ on  the output screen is determined by  the input field $E_n$, $n=1,\ldots, N$ and by the structure of  the transmission matrix $\mathbb{T}$  whose elements $\mathbb T_{n\nu}={t}^{(\nu)}_n$ connect input and output fields. The intensity transferred from all inputs to the $\nu$ output target is written as  
\begin{equation}
    \left\lvert E^{(\nu)} \right\rvert^2=\left\lvert \sum_n^N \tilde{t}^{(\nu)}_n E_n\right\rvert^2=
       \sum_{n,m} E_n t_n^{(\nu)} \bar t_m^{(\nu)} \bar E_m =
   \sum_{n,m} E_n {\mathbb {T}}_{n\nu} {\mathbb T}^\dag_{\nu m} \bar E_m 
   =\sum_{n,m} E_n {\mathfrak T}_{nm} \bar E_m 
\end{equation}
 where we have defined  the effective coupling matrix ${\mathfrak{T}}_{nm}={\mathbb {T}}_{n\nu} {\mathbb T}^\dag_{\nu m}$ between input modes $n$ and $m$
 and $\dag$ indicates the conjugate and transpose operation. For a single target mode, matrix $\mathbb T$ reduces to a vector $\mathbf{{t}}^{(\nu)}$ whose elements \textit{${t}^{(\nu)}_n$} encode attenuation and dephasing of a light rays incoming form the input modes \textit{n} to $\nu$. 
Since we now focus on one single target we will drop the output index $\nu$ in the following. In this single pattern case the coupling matrix reads 

\begin{equation}
\mathfrak{T}=\mathbf{{t}}\mathbf{{t}}^\dag.
\label{eq:CM}
\end{equation}
The elements of the  $\mathfrak{T}_{nm}$ matrix are not independently  randomly distributed. As it is resulting from the external product of a vector, $\mathbf{t}$, with itself   all columns and rows will depend on each other. The matrix has only one eigenvector.

In typical wavefront shaping experiments, the light  field $\bm E$ before the modulator is steady and, thus, it is convenient to define a vector $\mathbf{\Theta}$ whose elements are 
\begin{equation}
    \Theta_n={t}_n E_n .
    \label{def-Theta}
\end{equation} 
By  employing $N$ segments, the intensity on the output location at the far side of the disordered material is fully defined by the coupling matrix $\mathfrak{J}=\mathbf{\Theta} \mathbf{\Theta}^\dag$ whose elements $\mathfrak{J_{nm}}=\Theta_n \bar\Theta_m$ contain all the interference pairs. 

It is well known (see \textbf{S.I.} and Ref.   \cite{PhysRevLett.104.100601}) that the eigenvector of ${\mathbf{\Theta}} \mathbf{\Theta}^\dag$ corresponds to the input configurations providing the highest amount of intensity at the target.
As often the ${\mathbf{\Theta}}$ vector, and consequently  the coupling matrix $\mathfrak{J}$,  is inaccessible or extremely difficult to measure, light focusing behind disordered materials is usually achieved by a \textit{sequential optimization} (SO) protocol in which many random iterations (amplitude or phase changes on the input modes) are performed in order to maximize the intensity at a target.

A recent paper \cite{leonetti2021optical} demonstrated a correspondence between the intensity optimization through disordered materials and the physics of disordered systems. Indeed, when the SO algorithm is employed to maximize the intensity at several output targets symultaneously, its driving equation is identical  to the expression describing the energy of a  neural network, the Hopfield model for memory storage \cite{hopfield1982neural,PhysRevA.32.1007,PhysRevA.35.2293}.
When few targets are involved, in the statistical mechanical description, the transmission through the random optical  system corresponds  to the regime  of memory recovery in the Hopfield neural network.  We will, therefore, focus on this regime. Indeed, in the analogy, the transmission matrix elements linking each input to a single output correspond to a memory pattern stored in a neural network. And the input light mode fields $E_n$ play the role of a neuron activity. 

The intensity optimization corresponds, from a statistical mechanics point of view, to a zero temperature dynamics of the modes in a corrugated random landscape of mode configurations whose minima are the optimized solutions. Each random transmission matrix yields a different corrugation, i.e., a different energy landscape. This implies that any  small energy barrier around a non-optimal configuration incidentally encountered in the optimization procedure would not allow to optimize any further.
For instance, if the input fields are combined to provide $N$  variables only taking $\pm 1$ values,
the system relaxing towards equilibrium can be stuck in a non-optimal solution because of a barrier that could be overcome just  by a single sign switch of the field. 
We will refer to this sign switch as ``spin-flip'', since the $\pm 1$ variable is termed Ising spin in the statistical mechanical analogue. The system is, thus, stuck in a ``one-spin-flip-stable''  minimum corresponding to a non-optimal intensity.
In the following we will study how the presence of such metastable (shallow, one-spin-flip-stable) minima affect the SO.

\section*{Results}

To investigate the effect of dynamic arrest in the SO, we perform repeated sequential optimizations at the same location  starting from different input patterns, recording  the final configurations $\sigma_r$ of our phase modulator. Else said, we realize different replicated dynamics of the system relaxing, at $T=0$, in the same landscape. The index $r=1,\ldots, R$ is the replica index. We take $R=60$  replicas per location. These measurements have been made possible by employing the ``super-pixel method'' \cite{goorden2014superpixel} approach in similar fashion to what has been shown in paper \cite{leonetti2021optical} \textbf{(see S.I.)}, in which each segment of a Digital Micromirror Device (DMD) works as a phase modulator which can deliver  a 0 or $\pi$ phase prefactor (equivalent to multiply the injected intensity by  +1 or -1 ). Also, the same segment can be placed in the ``silent'' (off) configuration, delivering no light to the disordered system. This approach is very fast, enabling to acquire many replicas in a few minutes due to the high framerate of the DMD (up to 22 kHz).
Measuring many replicas enables to test the efficiency of the SO and, eventually, to expose the presence of possible sub-optimal states.

\begin{SCfigure*}[\sidecaptionrelwidth][t]
\centering
\includegraphics[width=11.4cm,height=11.4cm]{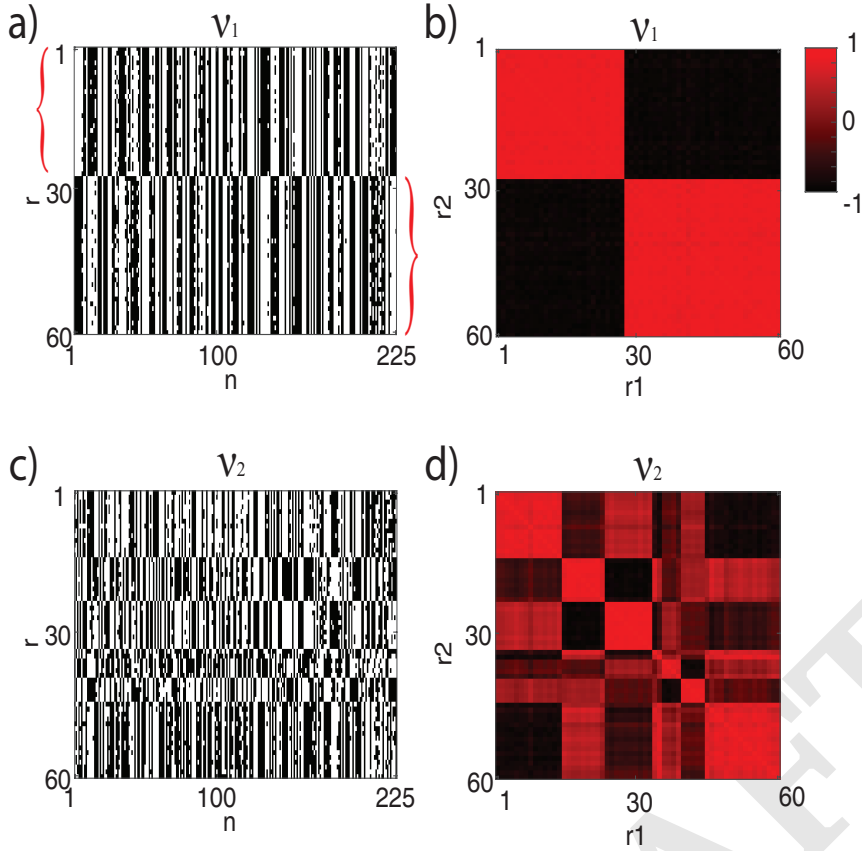}
\caption{ Panels a) and c) report the  configurations $\sigma_r$ obtained for different replicas of the experiments respectively in output targets $\nu_1$ and $\nu_2$. Note that the $r$ index has been sorted in order to have $\sigma_r$ pertaining to the same state clustered. a) shows two different clusters (the first from row 1 to row 28) the second 7 clusters (the first from row 1 to row 15). Panels b) and d) (respectively corresponding to $\nu_1$ and $\nu_2$) show the degree of similarity $Q$ calculated for all the replicas to all the replicas  (with  self-overlap reference $Q(\sigma_{r},\sigma_{r})=1$). The existence of multiple clusters is confirmed by this representation of $Q$.}
\label{fig: Sigma & Q}
\end{SCfigure*}

\begin{SCfigure*}
\centering
\includegraphics[width=.6\linewidth]{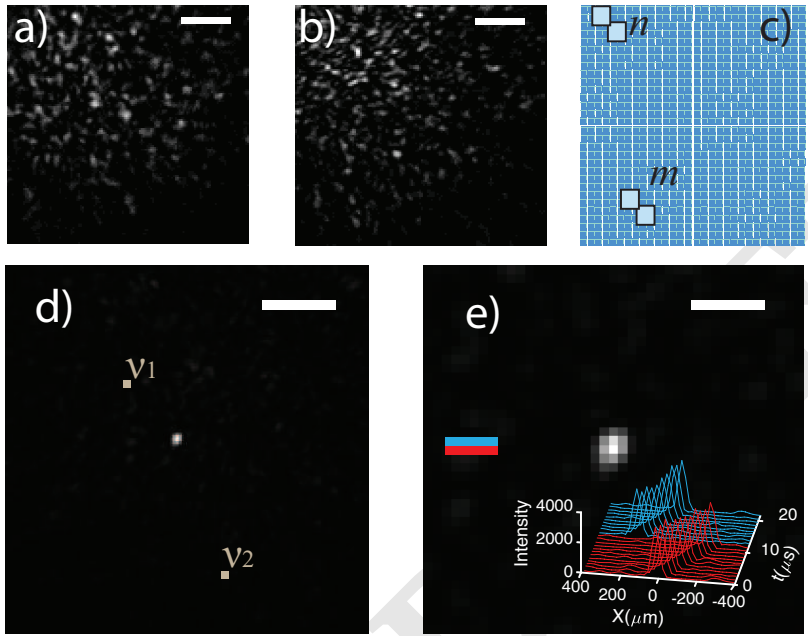}
    \caption{ Full TM measurements by CCM: panels a) and b) report the light intensity captured on the camera when only the mode \textit{n} is $+1$ (a) or when only the modes \textit{n} and \textit{m} simultaneously are in the $+1$ state (b), enabling to retrieve both $I_{n,n}$ and $I_{n,m}$. The sketch c) represent the configuration of the DMD corresponding to b). The small squares are the DMD micromirrors which are organized in 2x2 segments. Two of them ($n$ and $m$) are ``activated'' (each represented as two larger azure squares arranged in diagonal fashion). In panel d) we report a typical light focus obtained from Eigenvectors obtained by the CCM method. Panel e) is the same as d) with higher magnification: scale bars are 500 $\mu m$ (in a), b) and d) )  and 150 in e) $\mu m$ respectively Inset of panel e) reports the intensity profiles obtained at different times during the fast scanning trhough the disordered medium. The color indicates the scanning row as indicated in the pannel e) Note that the time axis in panel f has been constructed disregarding  sensor related delays (data transfer and camera readout). }
    \label{fig: DMD_SPM1}
\end{SCfigure*}

In Fig. \ref{fig: Sigma & Q} we report the (sorted) results for two, seemingly illustrative, different target locations ($\nu_1$ and $\nu_2$ also reported in Fig \ref{fig: DMD_SPM1}d)).

In particular, panels \ref{fig: Sigma & Q}a) and \ref{fig: Sigma & Q}b), refer to a location ($\nu_1$) displaying only two minima and \ref{fig: Sigma & Q}c) and \ref{fig: Sigma & Q}d),  refer to a location ($\nu_2$) with more (meta)stable configurations.  
Here the final states $\sigma_r$ are represented with black and white pixels corresponding to DMD segments placed in the -1 or +1 segment configuration, respectively.
In panel \ref{fig: Sigma & Q}a) the first $28$  $\sigma_r$ rows, (indicated by a red brace on the left) resulting from as many sequential optimization processes, are very similar, pertaining to a first cluster of measurements. On the other hand, rows $29$-$60$ (indicated by a red brace on the right) are, instead, pertaining to a second cluster of $\sigma_r$. Panel \ref{fig: Sigma & Q}b) reports the degree of similarity $Q$  between all the measurements acquired at target $\nu_1$.
The degree of similarity is defined as the scalar product between configurations: $$Q(\sigma_{r1},\sigma_{r2})=\frac{1}{N}\sigma_{r1}  \cdot \sigma_{r2}.$$ A value of $Q$ close to one indicate that  $\sigma_{r1}$ and $\sigma_{r2}$ are very similar, while $Q = 0$ indicates orthogonal input vectors.  $Q= -1$  indicates spin reversed vectors.

At target instance $\nu_1$  two different states can be clearly identified ($\sigma^\uparrow$ and $\sigma ^\downarrow$) and  they correspond to two ``spin reversed'' solutions ($\{\sigma_n\}\rightarrow \{-\sigma_n\}, ~\forall~n$). In other words, at location $\nu_1$ a single optimal configuration is found, together with its additive opposite. The intensity landscape, akin to the potential energy landscape in statistical mechanics, only displays a couple of global minima (degenerate and symmetric under spin inversion) making the optimum solution easy to recover.  Location $\nu_1$ has a number of clusters $N_C$ equal to $2$. 

The situation is quite different at location $\nu_2$ where multiple clusters are identified. The presence of many clusters is connected to the roughness  of the potential energy landscape of the system and the presence of shallow metastable minima and more than a single couple of stable minima. The instance at location $\nu_2$ turns out to display $N_C=7$.

Notice  that even if $\nu_1$ and $\nu_2$ are distant in space and, thus, are  identified by two completely different coupling matrices ${\mathfrak{T}}$, they have been selected for a fair comparison as they display very similar pre-optimization average intensity.

\begin{SCfigure*}
\centering
\includegraphics{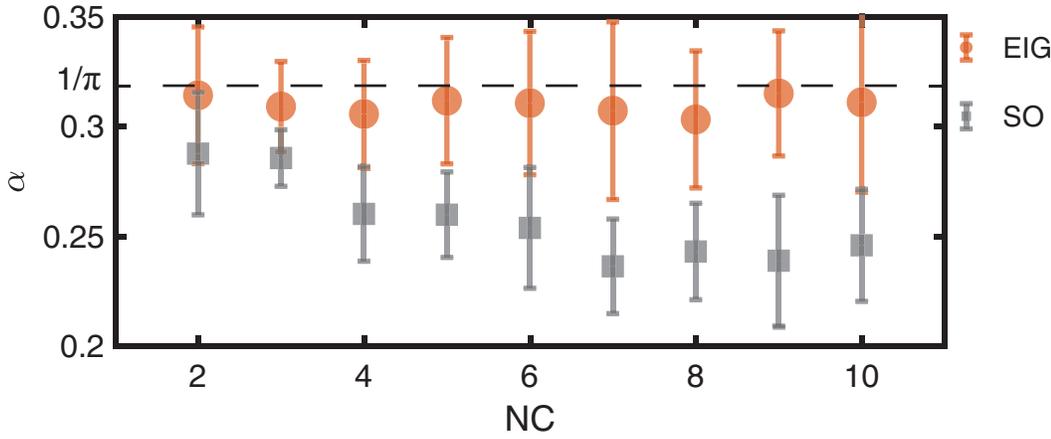}
\caption{The efficiency parameter $\alpha$ versus the number of clusters $NC$. Gray squares indicate results obtained with the SO approach while orange disks are obtained with the CCM. The dotted line represent the theoretical value for the average CCM approach. }
\label{fig: Alpha_NC}
\end{SCfigure*}

To understand the origin of such variability in the output, we performed the same experiment on many different output locations  $\nu$, namely $174$ targets, computing for each location the number of clusters $N_C$, that is a proxy  for the number of arrested, one-spin-flip-stable,  sub-optimal configurations in the system.  In all cases where $N_C>2$, indeed, taken away the possible optimal state (and its spin-reversed one), there will be  configurations of sub-optimal intensity.  

We also compute the focusing efficiency $\alpha$ defined in Eq. (\ref{eq:efficiency}). This is shown in Fig.\ref{fig: Alpha_NC}) (gray squares) plotted versus the total number of states $N_C$ where the error bars indicate the standard deviation of the $\alpha$ distribution. 
It is possible to note how the optimization efficiency $\alpha$ decreases when $N_C$ increases and how  $\alpha$ is always smaller than the theoretically predicted value  $1/\pi$  (indicated by a dashed line in the graph,  \textbf{see also S.I.}) of the maximal focusing efficiency.

To further investigate the role of the number of clusters, we resorted to a new approach to measure the  coupling  matrix Eq. (\ref{eq:CM}), the Complete Couplings Mapping Method  (CCMM), which is depicted in Fig. \ref{fig: DMD_SPM1}.

Indeed, by employing the DMD, we are able to tune each segment in the +1 or -1 or {\em off} (no light delivered to the experiment) configuration. 
Thus, by turning {\em off} all the segments except a pair $(n,m)$, we are able to estimate the individual elements of  $\mathfrak{J}$ without the need of any reference beam. To measure the full coupling matrix, we need to individually measure all the matrix elements, turning on sequentially all the $N$ segments and successively all the $N(N-1)/2$ pairs of segments. With $N=256$, e.g., we need to perform $32896$ measurements: a prohibitive task with a (slow) liquid crystal based  spatial light modulator, but perfectly feasible in few minutes with a fast DMD.

\begin{SCfigure*}
\centering
\includegraphics[width=.7\linewidth]{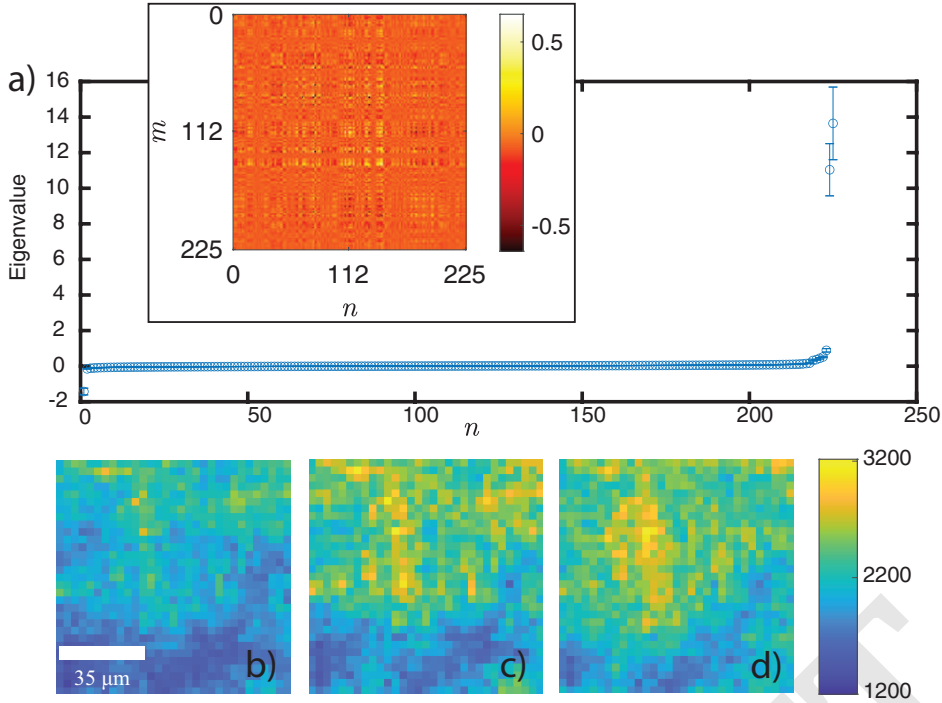}
\caption{a) Sorted  eigenvalues of the bi-dyadic single target coupling matrix associated to the TM. The inset shows the real coupling matrix  $J$ matrix: the values of $J_{nm}$ are reported in image format versus the $n$ and $m$ modulator segment index. The presence of highly valued rows or columns is resulting from the highly correlated nature of $J$  as it results from the external product $\mathbf{t} \mathbf{t}^\dag$. In panel b) we report the $I_{\rm focus}$ obtained with SO. Panel c) reports the $I_{\rm focus}$ estimated employing Eq. (\ref{Estimate of Ifocus}) while panel d) provides the $I_{\rm focus}$ obtained experimentally. All b), c) and d) matrices are retrieved in the same ROI. }
\label{fig: R matrix}
\end{SCfigure*}

By turning in the $+1$ configuration only in the input segments $n,m$, we retrieve the intensity $I_{nm}$ from which the transmission matrix can be extracted. In fact, if we express the elements  of the transmission matrix eigenvector (\ref{def-Theta}) in their real and imaginary parts,  $\Theta_n=\xi_n +\imath\chi_n$,  we obtain  (\textbf{see S.I.}):
\begin{eqnarray}
J_{nn}&=& I_{nn}
\label{intensity_JO_diag}
\\
    J_{nm}&=&J_{mn}=\frac{I_{nm}-I_{nn}-I_{mm}}{2}
        \label{intensity_JO_offdiag}
    \\
    \nonumber
   \mbox{where }  J&\equiv&\text{Real}(\mathfrak{J}).
\end{eqnarray}
We report the measured coupling matrix, estimated employing equations (\ref{intensity_JO_diag}) and  (\ref{intensity_JO_offdiag}) in the inset of  Fig. \ref{fig: R matrix}a. A couple of interference patterns and a sketch of a typical DMD configuration are reported in Fig. \ref{fig: DMD_SPM1}
In our experimental setup, intensity is recorded by a camera with a  Region Of Interest (ROI) of $128 \times 128$ pixels. Thus $P=16384$ different locations $\nu$ can be recorded simultaneously in a single measurement run.
Then, for each location $\nu$, the elements of the real part of the coupling matrix $\mathfrak{J}$ are  extracted from intensities, employing equations (\ref{intensity_JO_diag})-(\ref{intensity_JO_offdiag}) and the eigenvalues $\lambda^{(\nu)}$ and the eigenvectors $\mathbf e^{(\nu)}$ are retrieved. In this setting the measurement procedure is completed in about $5$ minutes. 
We realized a custom intensity analysis software which handles this extended data-set, retrieving all the elements of the coupling matrix. With the proposed formalism the ``open channels'' can be excited presenting as an input pattern for $J$ the eigenvector $\mathbf{e}$, see  \textbf{supplementary informations}.

Now we proceed to analyze the coupling  matrix $J$. It is a bi-dyadic real-valued matrix constructed with the Hebb rule starting from the real   and imaginary   coefficients of the transmission matrix from the whole input to a single target,  $\boldsymbol{\Theta}=\boldsymbol{\xi}+i\boldsymbol{\chi}$.
In particular, 
\begin{equation}
    J=\boldsymbol{\xi}\boldsymbol{\xi}^T+ \boldsymbol{\chi} \boldsymbol{\chi}^T, 
\end{equation}
the $\boldsymbol {^{T}}$ symbol indicating the transpose vector operation.
In other words $J$ is constructed with the Hebb rule starting from  two  vectors   $\boldsymbol{\xi}$ and $\boldsymbol{\chi}$ whose relationship is random and unknown.
This kind of interaction has been intensively studied in framework of neural networks \cite{amit1985storing,amit1985spinglass,amit1987statistical}, where synaptic connection matrices are constructed starting from the summation of dyadics in the framework of memory retrieval. In that framework when a ``stimulation'' pattern  corresponds to the eigenvector, then the corresponding memory state (open channel) is perfectly retrieved.
In the case of light transmission across random media the ``patterns'', i.e., the transmission matrix rows, are complex valued and, therefore, the coupling matrix through one target comes out to be bi-dyadic.
Bi-dyadic matrices possess at most two non-zero  eigenvalues \textbf{(see S.I.)}.
We notice that the more aligned $\boldsymbol \xi$
and $\boldsymbol \chi$ are,  the more degenerate  the non-zero eigenvalues will be. Indeed, if $\boldsymbol\xi = \boldsymbol\chi$ the matrix $J$ is a simple dyadic, with a single non-zero eigenvalue e a single eigenvector, corresponding to a single open channel.


The average structure of the eigenvalues found from experimental data, is reported in Fig. \ref{fig: R matrix}a). We measure two positive eigenvalues  (the 224 $th$ and the 225 $th$) and a set of $N-2$ eigenvalues very close to 0, as expected for a bi-dyadic matrix. A small deviation form the ideal picture is due to noise in the interference measurement.

We can employ the most intense   eigenvector $\mathbf{e}^{\rm max}$, related to the largest eigenvalue, to transmit through the most transmitting open channel. Eigenvectors are composed by $N$ real numbers, while the DMD is only capable to deliver $N$  values  $\pm 1$ in input. Thus, to get the DMD input array which is best approximating the eigenvector $\mathbf{e}^{\rm max}$ elements  we define the vector of signs $\boldsymbol{\sigma^{\rm max}}=\text{sign}(\mathbf{e}^{\rm max})$. 
The resulting modulation factor $\alpha$ obtained within this approach is reported in Fig.  \ref{fig: Alpha_NC}) as orange circles versus the number of clusters $N_C$.


It is possible to note that $\alpha$ obtained in this case is larger than the one obtained with sequential optimization, it does not depend on $N_C$ and it is very  close  to the theoretically expected value of $\alpha_{theor}=1/\pi \simeq 0.31831 $ \textbf{See S.I.}.
The CCMM, indeed, provides 
$\alpha_{exp}=0.3125\pm 0.0021$, which is $98 \%$ of the theoretical value (estimate obtained averaging the values from in Fig. \ref{fig: Alpha_NC} for all $NC$ ).  
We stress that the CCMM estimate of $\alpha$ is independent of the number of clusters $N_C$ occurring at the various target positions $\nu$, as it was expected since the  approach is not affected by the roughness of the intensity landscape. Note that  previous methods to experimentally determine $\mathbf t$, related to the eigenvector of  $J$, are sub-optimal \textbf{(see supplementary information)}, reaching from 40$\%$ to 60$\%$ of the theoretically expected $\alpha$, typically because they are strongly influenced by the reference patterns \cite{PhysRevLett.104.100601}.
Lower panels of Fig. \ref{fig: R matrix}, report the intensity obtained at the end of the optimization process with SO (b) and with the eigenvector found by CCMM  (d). The two pictures are similar but the eigenvalue approach based on CCMM provides optimized intensity always higher than  SO.
Fig. \ref{fig: R matrix}c reports the predicted intensity obtained by extracting the eigenvalue and   the  eigenvector  $\boldsymbol{\sigma}^{\rm max}$ from the measured  coupling matrix and then using the formula  
\begin{equation}
    I_{\rm opt}^{(\nu)}=\sum_{nm} \sigma^{\rm max}_n\,    J_{nm}\, \sigma^{\rm max}_m.\label{Estimate of Ifocus}
\end{equation}

\begin{SCfigure*}
\centering
\includegraphics[width=.7\linewidth]{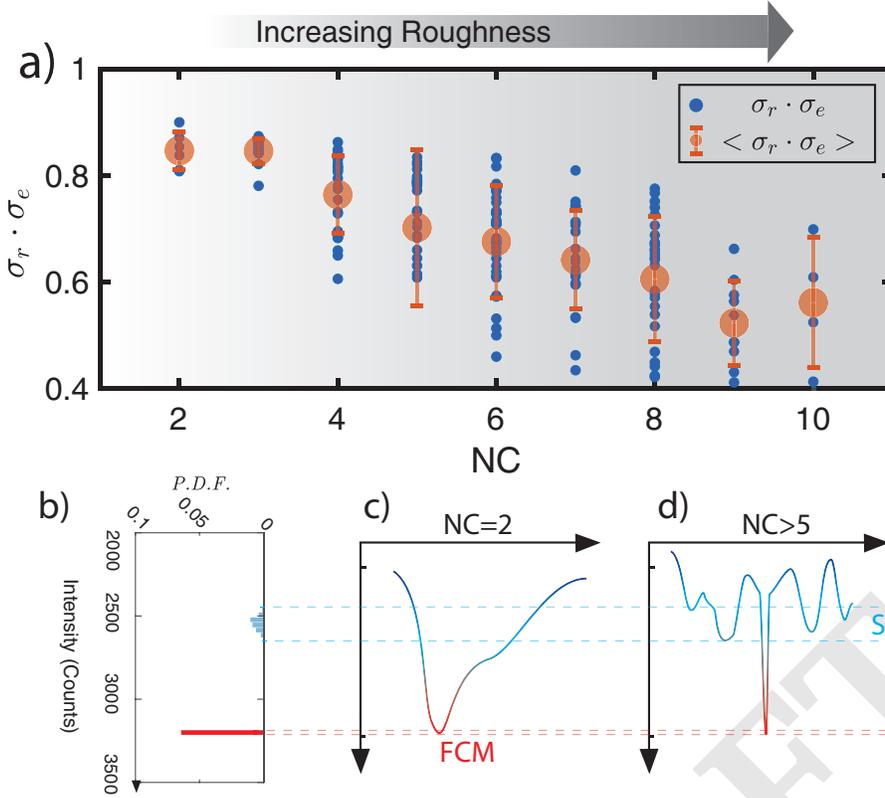}
\caption{a) reports the overlap $\boldsymbol \sigma^{(r)} \cdot \boldsymbol \sigma^{\rm max}$ versus $N_C$, $\boldsymbol \sigma^{(r)}$ being the SO configuration, $\boldsymbol \sigma^{\rm max}$ being the eigenvector of the retrieved transmission channel by means of CCMM. The orange circles are the average values (error bars are the statistical error), and smaller blue dots are individual measurements.  Panel b) reports the  probability density function of the intensity ($N$=225 at a single location) for SO (azure) and CCMM (orange).  The hand drawn pictures in panels c) and d) report a qualitative view of how could be imaged the Intensify landscape for the low NC and for the high NC case.  }
\label{fig: OV Vs NC}
\end{SCfigure*}

To further investigate the relation between roughness of the intensity landscape and focusing efficiency $\alpha$ obtained withi the SO we plot in Fig. \ref{fig: OV Vs NC} the overlap (degree of similarity)  calculated with the normalized scalar product  $\boldsymbol\sigma^{(r)} \cdot  \boldsymbol{\sigma^{\rm max}}$ between the  optimized input vector resulting  from a sequential optimization replica $\boldsymbol{\sigma}^{(r)}$ and the input profile obtained by eigenvalue retrieval $\boldsymbol{\sigma^{\rm max}}$.

It is possible to note how higher $N_C$ are correlated with sequential optimizations far from the optimal solution $\boldsymbol\sigma^{\rm max}$. A sketch of the potential energy landscape representing the two configuration is shown in the bottom sketches. Fig. \ref{fig: OV Vs NC}b) reports the probability density function of the final intensity, obtained for the CCMM (red) and for the SO. The graph demonstrates that the CCMM retrieves a much higher intensity ($+28 \%$ over the field shown in Fig. \ref{fig: R matrix}(b-c)) and provides a smaller intensity variance. The hand-drawn pictures of the landscape generating this results is reported in Fig. \ref{fig: OV Vs NC}c-d)

\section*{Acknowledgments}

We  acknowledge Prof. Giorgio Parisi for scientific exchange, the support from the European Research Council (ERC) under the European Union’s Horizon 2020 Research and Innovation Program, Project LoTGlasSy (Grant Agreement No. 694925), and  
 the support of LazioInnova - Regione Lazio under the program {\em Gruppi di ricerca 2020} - POR FESR Lazio 2014-2020, Project NanoProbe (Application code A0375-2020-36761) and  Project LocalSscent, (Application code. A0375-2020-36549).

\providecommand{\noopsort}[1]{}\providecommand{\singleletter}[1]{#1}%

\end{document}